\documentclass{optica-article}
\usepackage{tabularx}
\usepackage{array}
\newcolumntype{Y}{>{\raggedright\arraybackslash}X}
\journal{opticajournal}

\articletype{Research Article}

\usepackage{lineno}

\usepackage{orcidlink}
\usepackage{amsmath,amssymb}
\usepackage{siunitx}
\usepackage{multirow}
\usepackage{booktabs}
\usepackage{graphicx}
\usepackage{subcaption}
\usepackage{xcolor}
\newcommand{\rev}[1]{\textcolor{blue}{#1}}

\begin{document}
\pagestyle{plain}
\pagenumbering{arabic}

\title{Inverse Design\textendash Validated Optimization of Lead-Free Cs$_3$Cu$_2$Cl$_5$ Visible-Light Microring Resonators via Coupled DFT-FDTD Framework}

\author{Shoumik Debnath,\,\orcidlink{0009-0001-9500-5319}\authormark{1} and Sudipta Saha\,\orcidlink{0009-0003-9110-8059}\authormark{1,2*}}

\address{\authormark{1}Department of Electrical and Electronic Engineering, Bangladesh University of Engineering and Technology, Dhaka 1000, Bangladesh\\
\authormark{2}Accident Research Institute, Bangladesh University of Engineering and Technology, Dhaka 1000, Bangladesh}

\email{\authormark{*}sudiptasaha@ari.buet.ac.bd}

\begin{abstract*}
Microring resonators are indispensable for wavelength filtering, sensing, and on-chip signal routing in photonic integrated circuits, yet visible-wavelength implementations using environmentally benign materials remain scarce.
We report a numerical design study of add-drop MRRs employing cesium copper chloride (Cs$_3$Cu$_2$Cl$_5$), a lead-free all-inorganic halide with favorable visible-band optical properties.
Wavelength-resolved refractive index ($n$) and extinction coefficient ($k$) of Cs$_3$Cu$_2$Cl$_5$, computed via density functional theory (DFT), serve as direct input to \rev{finite-difference time-domain (FDTD) modeling: 2.5D variational FDTD (varFDTD) for the parametric sweeps, and full three-dimensional FDTD for validation of the selected geometry}.
Independent parametric scans cover ring waveguide width (500--900~nm), coupling gap (150--300~nm), and bend radius (5--20~$\mu$m).
At the balanced operating point of 600~nm ring width, 200~nm gap, and 10~$\mu$m radius, we obtain a loaded quality factor ($Q$) of ${\sim}5{,}386$, a free spectral range (FSR) of 11.3~nm, a drop-port extinction ratio (ER) of 32.2~dB, and a finesse of 95.8.
The gap sweep traces the full trajectory from over-coupled through critically coupled to under-coupled operation, with the critical point near 200~nm.
A sharp bending-loss cliff emerges between 5 and 10~$\mu$m, below which all figures of merit collapse.
These sweeps yield a systematic geometry--performance map for Cs$_3$Cu$_2$Cl$_5$-based MRRs\rev{, the first reported for this compound to our knowledge}. Cross-platform validation in Tidy3D reproduces the spectral fingerprint of the optimized device, and inverse design of the bus coupling region delivers a 3\% gain in drop-port power transfer, indicating that the sweep-derived geometry sits close to the coupling optimum.
\end{abstract*}

\section{Introduction}

Microring resonators (MRRs), closed-loop optical cavities evanescently coupled to bus waveguides, have become versatile functional elements in photonic integrated circuits \cite{Mirza2018MRR}. Light circulating inside the ring interferes constructively at discrete wavelengths, producing narrow spectral features exploitable for channel selection, label-free biosensing, and reconfigurable switching \cite{Bogaerts2012, Rabus2007}. Adding a second bus creates the add-drop topology. Resonant light exits via the drop port while off-resonant wavelengths continue to the through port \cite{Little1997}. This arrangement underpins wavelength-division demultiplexing, modulation, and $n$ transduction \cite{Bogaerts2012, Chrostowski2015}. Performance is benchmarked by the loaded quality factor ($Q$), \rev{the free spectral range (FSR), the extinction ratio (ER),} and finesse ($\mathcal{F}$) \cite{Bogaerts2012, Yariv2000}. 

Most MRR research to date has targeted silicon-on-insulator (SOI) at telecom wavelengths ($Q > 10^5$, few-$\mu$m radii) \cite{Bogaerts2012}, stoichiometric Si$_3$N$_4$ for visible and near-infrared operation \cite{Subramanian2013, Morin2021}, or thin-film lithium niobate (TFLN) for electro-optic and nonlinear functionality \cite{Zhu2021, Zhang2017LN}. The vast majority of demonstrated devices operate in the near-infrared \cite{Haas2019RSC}. There is, however, growing motivation to push MRR technology into the visible for biological sensing, Raman spectroscopy, and quantum light sources \cite{Morin2021, Tran2022}. 

Organic--inorganic lead halide perovskites CH$_3$NH$_3$PbX$_3$ (X\,=\,Cl, Br, I) exhibit outstanding optoelectronic figures of merit \cite{Kojima2009, Green2014}, but soluble lead raises toxicological and regulatory concerns \cite{Li2020Nat, Babayigit2016}. This has spurred an active search for lead-free substitutes \cite{Luo2020, Lian2020}. 

The ternary copper halide Cs$_3$Cu$_2$Cl$_5$ is one such candidate. It exhibits bright green self-trapped-exciton emission, photoluminescence quantum yields near unity, and robust thermal stability, all from earth-abundant, low-toxicity constituents \cite{Ali2023Heliyon, Zhang2020}. Its zero-dimensional crystal lattice consists of discrete [Cu$_2$Cl$_5$]$^{3-}$ clusters separated by Cs$^+$ cations, and the resulting strong electron--phonon coupling localizes excitons and suppresses non-radiative decay \cite{Zhang2020, Luo2020}. The material has been deployed in thin-film LEDs \cite{Wang2020LED}, X-ray scintillators \cite{Zhou2022}, and photodetectors \cite{Cheng2019}. Its use as a passive waveguide core for photonic resonators, however, has not been investigated. 

Visible-band MRR demonstrations remain scarce and largely confined to Si$_3$N$_4$ \cite{Subramanian2013, Morin2021}. No geometry-dependent performance map exists for Cs$_3$Cu$_2$Cl$_5$-clad resonators. Existing literature on this compound addresses bulk or thin-film luminescence rather than guided-wave device physics. Constructing a geometry-dependent performance map is a prerequisite for translating the material's favorable optical window into functional photonic components. 

Inverse design, where device geometry is treated as a set of continuous optimization variables updated via gradient information, has reshaped photonic component engineering \cite{Molesky2018, Jensen2011}. Applied to an MRR coupling section, it can quantify how far a parametric sweep result sits from the true optimum. No such study has been reported for lead-free halide platforms at visible wavelengths.

\rev{Studies of halide perovskites in integrated photonics have concentrated on active functionality, with the perovskite serving as an emissive or nonlinear overlay on a conventional guiding platform: Wang \emph{et al.}\ exploited the optical nonlinearity of a metal-halide perovskite to control the extinction ratio of a microring resonator \cite{Wang2021AOM}, and Han \emph{et al.}\ reported inorganic-perovskite active multifunctional photonic devices \cite{Han2024NC}; in both, the perovskite supplies an active optical function rather than acting as the passive guiding medium. Studies of Cs$_3$Cu$_2$Cl$_5$ itself have been material-level rather than device-level, addressing electronic structure, optical constants and luminescence in bulk and thin-film form \cite{Ali2023Heliyon, Zhang2020, Zhou2022}. Computational optimization of microring geometry is well established; Dizaji and Habibiyan, for instance, coupled Bayesian optimization to physics-based constraints to select microring geometry for squeezed-light generation \cite{Dizaji2025SR}. What has not been done, to our knowledge, is to apply this optimization method to a lead-free copper halide acting as the guiding medium itself. What distinguishes the present study is the combination: a lead-free copper halide treated as the passive guiding core, with optical constants taken from first principles rather than from fitted dispersion models, mapped across a three-parameter geometry space in the visible band and then checked against an independent solver and a gradient-based optimization of the coupling region.}

In this paper we couple DFT-derived optical constants with full-wave FDTD modeling to chart the design landscape of Cs$_3$Cu$_2$Cl$_5$ add-drop MRRs. \rev{The refractive index $n$ and the extinction coefficient $k$ are obtained from DFT and supplied to the solver as a spectrally resolved material model.} Systematic scans of ring width (500--900~nm), coupling gap (150--300~nm), and ring radius (5--20~$\mu$m) are performed in Lumerical. The resulting spectra yield $Q$, FSR, ER, $\mathcal{F}$, and insertion loss (IL). The best-performing geometry delivers $Q \approx 5{,}400$. An independent Tidy3D simulation reproduces the spectral response, and inverse design of the coupling region supports the near-optimality of the parametric result. These outcomes provide a device-level design ruleset for lead-free, visible-regime MRRs based on Cs$_3$Cu$_2$Cl$_5$\rev{, which to our knowledge has not previously been reported}.

\rev{The rest of this paper is organized as follows. Section~3.1 examines the effect of ring width on resonator performance, followed by the coupling-gap and ring-radius analyses in Sections~3.2 and~3.3, respectively. Section~3.4 combines these results to identify the optimized geometry, while Section~3.5 discusses the physical limits of the resulting design space. Section~3.6 analyzes the supported modes and modal confinement. Section~3.7 validates the selected geometry using an independent three-dimensional FDTD solver and further refines the coupling region through gradient-based inverse design. Finally, Section~3.8 evaluates fabrication tolerance and practical feasibility.}

\section{Device design methodology}

\subsection{DFT calculations for optical constants}

\begin{figure}[htbp]
\centering
\begin{subfigure}[t]{0.45\linewidth}
    \includegraphics[width=\linewidth]{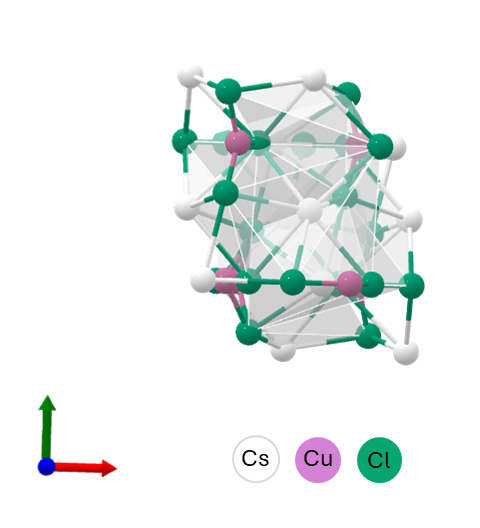}
    \caption{}
\end{subfigure}
\hfill
\begin{subfigure}[t]{0.5\linewidth}
    \includegraphics[width=\linewidth]{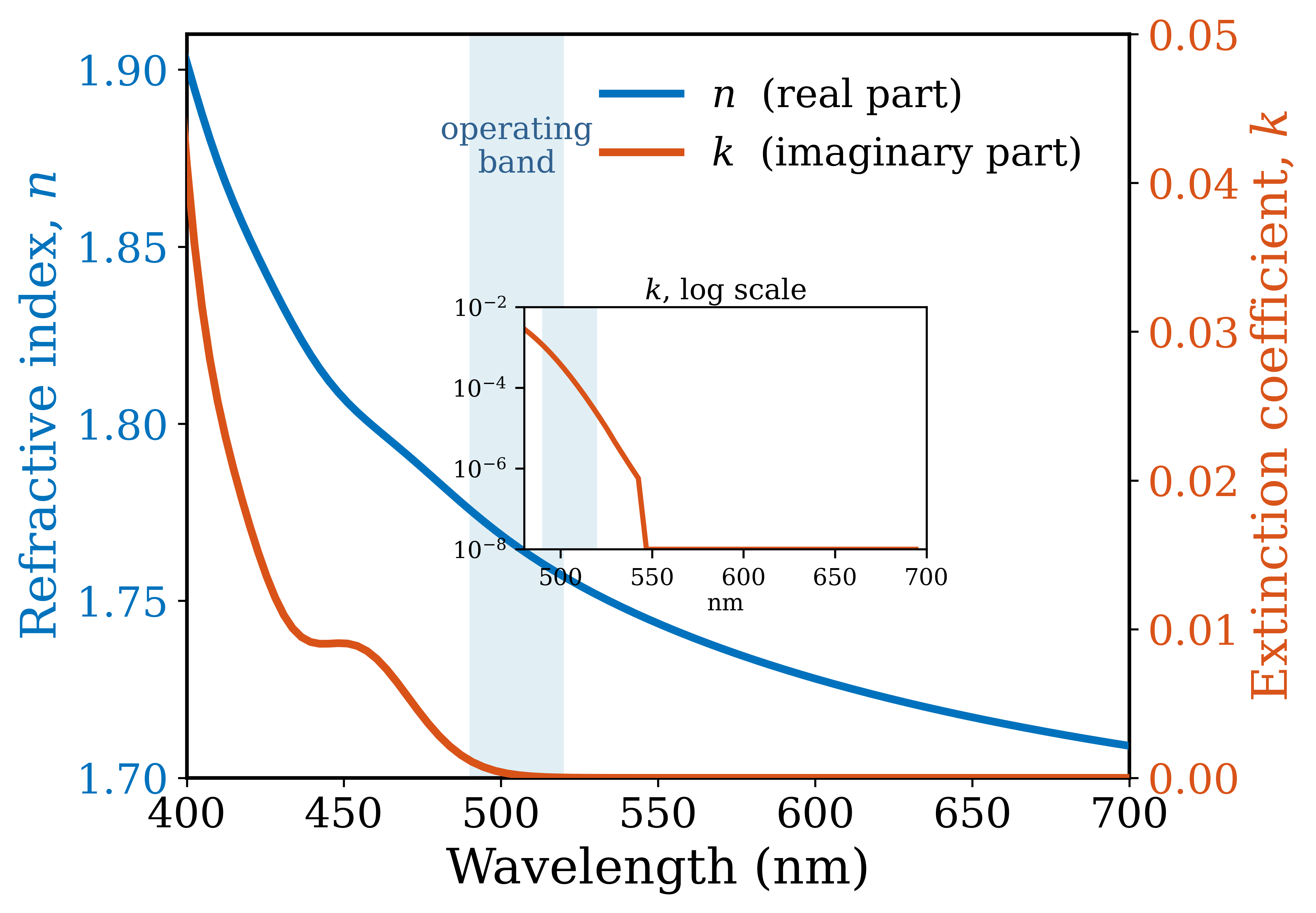}
    \caption{}
\end{subfigure}
\caption{(a) Unit cell of Cs$_3$Cu$_2$Cl$_5$ ($Pnma$), rendered from Materials Project data (mp-582024) \cite{Jain2013}. (b) DFT-computed $n$ and $k$ across the visible spectrum. The shaded region indicates the primary operating wavelength range.}
\label{fig:nk_plot}
\end{figure}

The structural compound of  Cs$_3$Cu$_2$Cl$_5$  crystallizes in the orthorhombic space group $Pnma$ and features a zero-dimensional motif in which discrete [Cu$_2$Cl$_5$]$^{3-}$ polyanions are embedded in a Cs$^+$ matrix (Fig.~\ref{fig:nk_plot}(a)). The structural model is adapted from the Materials Project entry mp-582024 \cite{Jain2013, Zhang2020, Luo2020}.

First-principles calculations were carried out with the CASTEP plane-wave code within Materials Studio. Atomic positions and lattice constants were relaxed under the \rev{generalized gradient approximation (GGA)} using the Perdew--Burke--Ernzerhof (PBE) functional until residual forces fell below 0.01~eV\,\AA$^{-1}$ and the total-energy change between successive steps was less than $10^{-6}$~eV. Because semilocal functionals systematically underestimate semiconductor band gaps, electronic band structure and optical spectra were subsequently computed with the HSE06 hybrid functional, with spin--orbit coupling (SOC) included. Ultrasoft pseudopotentials, a 500~eV plane-wave cutoff, and a $4 \times 4 \times 2$ Monkhorst--Pack $k$-mesh were used throughout. The frequency-dependent complex dielectric tensor was evaluated from interband transition matrix elements. Its real and imaginary parts were then converted into wavelength-dependent $n(\lambda)$ and $k(\lambda)$ over the 400--700~nm window.

\rev{The refractive index $n$ and the extinction coefficient $k$ used throughout this work are the real and imaginary parts of the complex refractive index $\tilde{n} = n + \mathrm{i}k$, and both are plotted in Fig.~\ref{fig:nk_plot}(b). They are related to the complex dielectric function $\varepsilon = \varepsilon_1 + \mathrm{i}\varepsilon_2$ computed directly by CASTEP through}
\begin{equation}
\rev{\varepsilon_1 = n^2 - k^2, \qquad \varepsilon_2 = 2nk,}
\label{eq:eps}
\end{equation}
\rev{so the dielectric representation can be recovered from the plotted curves without recomputation.}

\rev{The geometry optimization and the subsequent optical calculation are static ground-state calculations carried out at zero applied pressure and without a thermostat. The optical constants therefore correspond to the relaxed ground-state structure at zero applied pressure and do not include explicit finite-temperature effects. The consequences of this for the device design are discussed in Section~3.5.}

Figure~\ref{fig:nk_plot}(b) presents the computed dispersion. $n$ falls from ${\sim}1.91$ near 400~nm to ${\sim}1.71$ at 700~nm, consistent with normal dispersion. $k$ stays below 0.01 for $\lambda > 500$~nm, \rev{indicating} a broad low-loss operating window. The shaded region marks the primary operating range (490--520~nm). These values were imported into Lumerical as a sampled-data material with piecewise-linear interpolation.

\subsection{Device structure}

\begin{figure}[t]
\centering
\begin{subfigure}[t]{0.62\linewidth}
    \includegraphics[width=\linewidth]{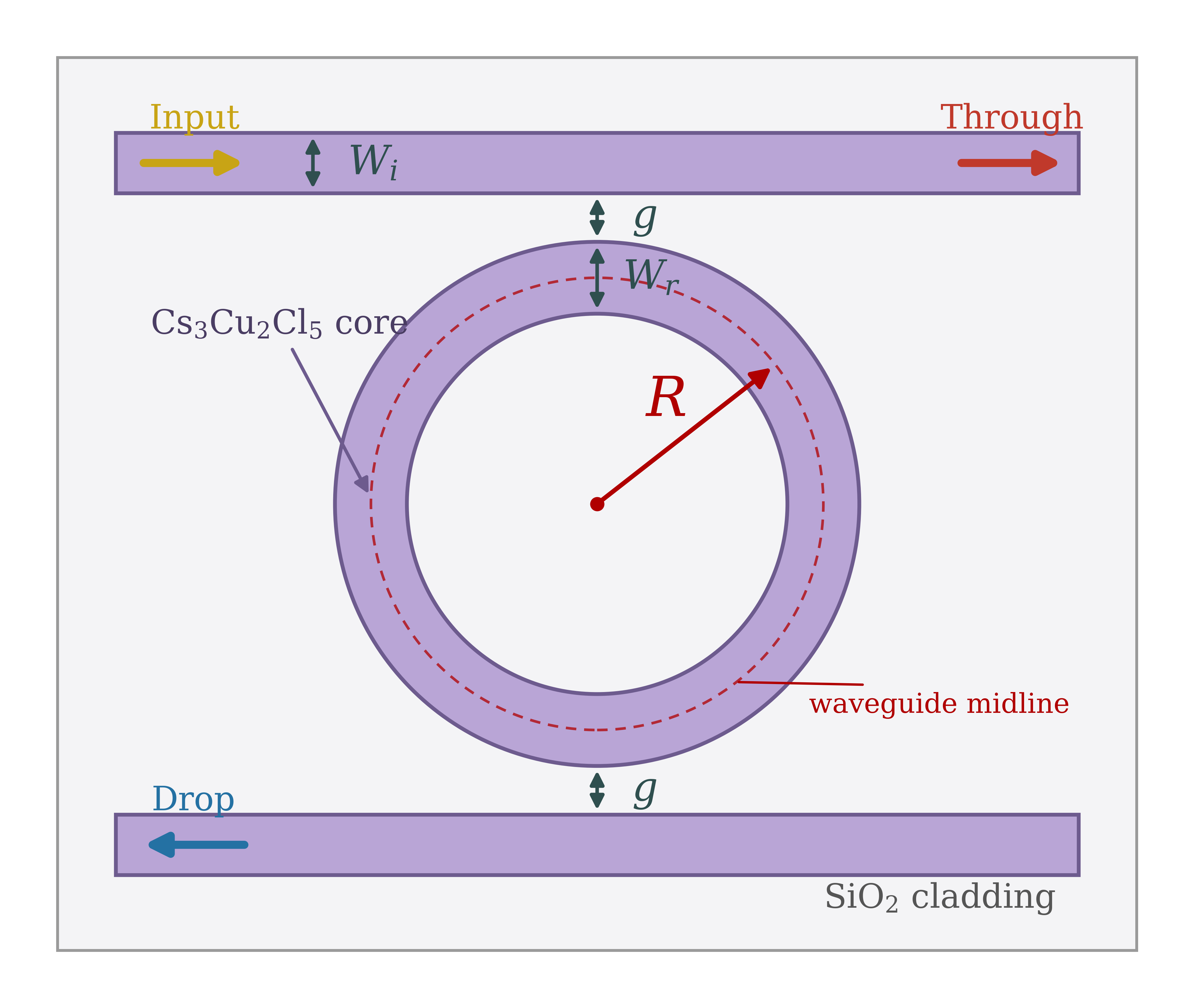}
    \caption{}
\end{subfigure}\\[6pt]
\begin{subfigure}[t]{0.92\linewidth}
    \includegraphics[width=\linewidth]{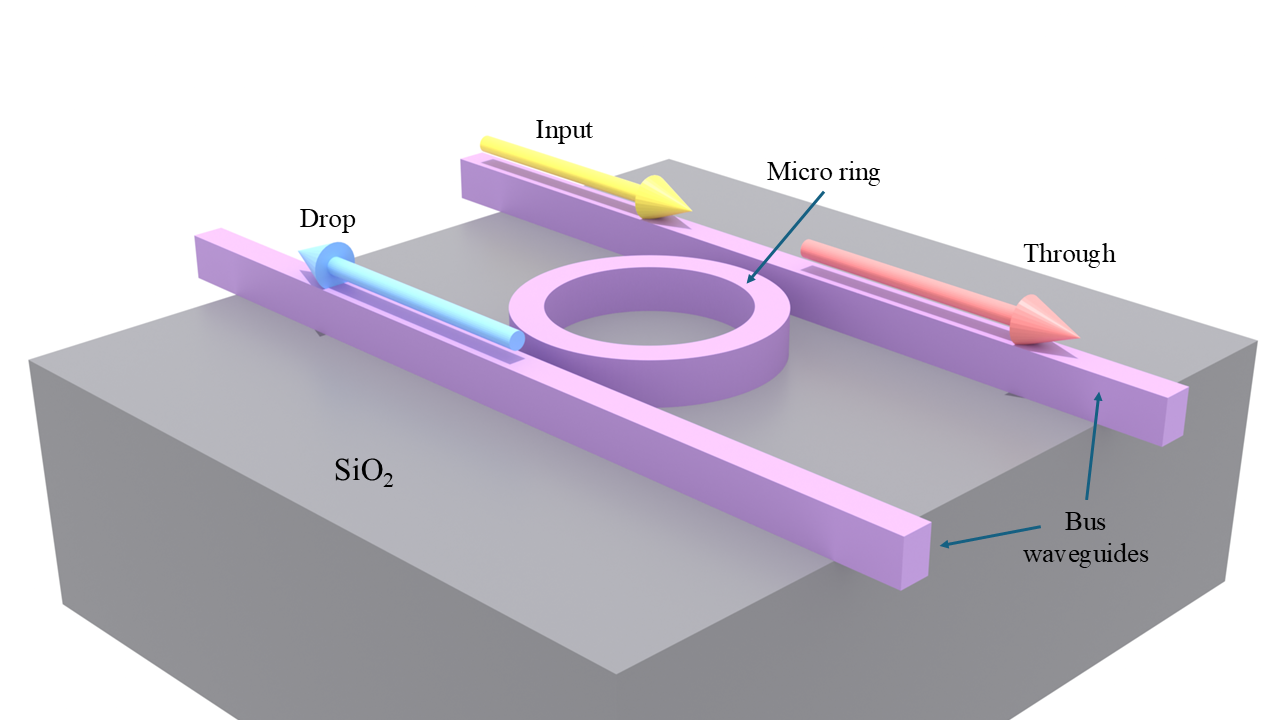}
    \caption{}
\end{subfigure}
\caption{(a) Plan-view layout of the Cs$_3$Cu$_2$Cl$_5$ add-drop MRR with labeled geometric parameters\rev{: ring width $W_r$, bus width $W_i$, coupling gap $g$, and ring radius $R$ measured from the ring center to the midline of the ring waveguide}. (b) Perspective rendering on SiO$_2$ substrate.}
\label{fig:schematic}
\end{figure}

The target device is a symmetric add-drop MRR in which a closed-loop waveguide sits equidistant from two parallel bus waveguides (Fig.~\ref{fig:schematic}). On resonance, input power transfers into the ring and exits through the drop port. Off-resonant wavelengths propagate to the through port \cite{Bogaerts2012, Rabus2007}.

Three geometric degrees of freedom are investigated\rev{: the ring waveguide width $W_r$, the coupling gap $g$, and the ring radius $R$}. \rev{The ring waveguide width $W_r$} is scanned from 500 to 900~nm in 100~nm increments. The coupling gap~$g$ is varied from 150 to 300~nm in 25~nm steps. \rev{The ring radius} $R$, measured from the ring center to the midline of the ring waveguide, takes the values 5, 10, 15, and 20~$\mu$m. Two parameters are held constant throughout. The bus width $W_i = 500$~nm and the film thickness $h = 400$~nm. Both ring and bus cores are Cs$_3$Cu$_2$Cl$_5$; the under-cladding and top cladding are SiO$_2$. To isolate the influence of each variable, only one parameter is swept at a time. The others remain at baseline values $W_r = 600$~nm, $g = 200$~nm, and $R = 10$~$\mu$m.

\rev{The width sweep requires attention, because widening the ring displaces its outer edge and would otherwise alter the coupling gap at the same time. The ring waveguide is therefore widened symmetrically about its midline, so that $R$, which is defined to the midline, remains fixed at 10~$\mu$m across the sweep; and the two bus waveguides are repositioned at each step so that the edge-to-edge separation between bus and ring remains exactly 200~nm. The width sweep consequently varies $W_r$ at constant radius and constant physical gap, and the trends reported in Section~3.1 are not contaminated by a drifting coupling condition. In the radius sweep, $W_r$ and $g$ are likewise held at their baseline values, with the bus waveguides repositioned to follow the ring; in the gap sweep, the ring is fixed and only the bus offset is changed. The bus width $W_i = 500$~nm throughout.}

\subsection{FDTD simulation setup}

\rev{Electromagnetic modeling proceeds in two tiers of fidelity. The parametric sweeps, which cover sixteen distinct geometries, are carried out with the 2.5D variational finite-difference time-domain (varFDTD) solver of Ansys Lumerical \cite{Lumerical, Taflove2005}; the selected geometry is then re-solved with full three-dimensional FDTD in an independent code (Section~3.7). The varFDTD method collapses the vertical dimension of a planar structure into an effective index and solves the resulting in-plane problem, which is what makes a sweep of this size tractable. It is appropriate here because the 400~nm Cs$_3$Cu$_2$Cl$_5$ film supports a single guided vertical mode over the wavelengths considered, which the cross-sectional eigenmode analysis reported in Section~3.6 supports, and because the quantities of interest are governed by in-plane geometry. The 3D calculation is used to verify that the selected optimum reproduces the principal spectral characteristics identified by the 2.5D sweep. The 2.5D results are accordingly presented as a comparative screening of the parameter space rather than as absolute predictions at every point, and we do not claim that the 3D solver independently reproduces the full sixteen-geometry map.}

A broadband eigenmode source injecting the fundamental TE$_0$ mode spanned 500--700~nm, placing several successive ring resonances within the simulation bandwidth for direct FSR extraction. Frequency-domain monitors at the through and drop facets recorded transmission coefficients.

\rev{The in-plane domain boundaries used perfectly matched layers (PMLs) \cite{Berenger1994}. Mesh refinement used Lumerical's conformal mesh technology under Conformal Variant 1, with a minimum cell of 5~nm globally and a finer override region spanning the coupling gaps. Conformal mesh technology resolves a dielectric interface lying within a Yee cell by assigning an effective permittivity weighted by the fraction of the cell each material occupies, rather than displacing the interface to the nearest grid point as staircasing does; this matters here because the coupling gap and the ring sidewall are both dielectric interfaces that do not in general align with the grid. We note that Conformal Variants 0, 1 and 2 differ from one another only in their treatment of interfaces involving metals or perfect electrical conductors, of which the Cs$_3$Cu$_2$Cl$_5$/SiO$_2$ structure contains none, so the three settings coincide for this geometry. Temporal integration continued until the residual field energy fell to $10^{-5}$ of its peak or until the iteration limit was reached, whichever occurred first. Table~\ref{tab:solver} collects the configuration of both tiers.}

\begin{table}[htbp]
\centering
\caption{\rev{Solver configuration for the two fidelity tiers.}}
\label{tab:solver}
\begin{tabularx}{\linewidth}{@{}l>{\raggedright\arraybackslash}X>{\raggedright\arraybackslash}X@{}}
\toprule
\rev{\textbf{Setting}} & \rev{\textbf{Parametric sweeps}} & \rev{\textbf{Validation / inverse design}} \\
\midrule
\rev{Solver}            & \rev{Lumerical 2.5D varFDTD}      & \rev{Tidy3D, full 3D FDTD} \\
\rev{Geometries}        & \rev{16 (width, gap, radius)}     & \rev{1 (selected optimum)} \\
\rev{Source}            & \rev{Broadband TE$_0$ eigenmode, 500--700~nm} & \rev{Broadband mode source, 620--650~nm} \\
\rev{Mesh refinement}   & \rev{Conformal mesh technology, Variant 1} & \rev{Automatic, 10 steps per wavelength} \\
\rev{Minimum mesh step} & \rev{5~nm, with override across the coupling gaps} & \rev{Set by steps-per-wavelength criterion} \\
\rev{Boundaries}        & \rev{PML on the in-plane domain boundaries} & \rev{PML on all six boundaries} \\
\rev{Material model}    & \rev{Sampled DFT $n(\lambda)$, $k(\lambda)$, piecewise-linear} & \rev{Non-dispersive, $n=1.720$ core, $n=1.457$ cladding} \\
\rev{Shutoff criterion} & \rev{$10^{-5}$ residual field energy} & \rev{$10^{-5}$ residual field energy} \\
\bottomrule
\end{tabularx}
\end{table}

\subsection{Performance metrics}

Five figures of merit are extracted from the simulated port spectra.

\emph{Loaded $Q$.} Ratio of resonance wavelength to $-3$~dB linewidth,
\begin{equation}
Q = \frac{\lambda_\text{res}}{\Delta\lambda_\text{FWHM}}.
\label{eq:Q}
\end{equation}

\emph{FSR.} Spacing between adjacent resonances of the same azimuthal order,
\begin{equation}
\text{FSR} = \frac{\lambda^2}{n_g \cdot 2\pi R},
\label{eq:FSR}
\end{equation}
with $n_g$ the group index. In practice, FSR values were read directly from the spectra.

\emph{ER.} On-to-off resonance contrast at each port, in decibels,
\begin{equation}
\text{ER} = -10 \log_{10} \!\left( \frac{T_\text{min}}{T_\text{max}} \right).
\label{eq:ER}
\end{equation}
\rev{Here $T_\text{min}$ and $T_\text{max}$ are the minimum and maximum transmission of the port under consideration, evaluated over one free spectral range about the resonance. At the through port the resonance appears as a minimum, so $T_\text{min}$ is the on-resonance value and $T_\text{max}$ the off-resonance baseline; at the drop port the resonance appears as a maximum, and the two are interchanged.}

\emph{Finesse.} Resolvable linewidths per FSR,
\begin{equation}
\mathcal{F} = \frac{\text{FSR}}{\Delta\lambda_\text{FWHM}}.
\label{eq:finesse}
\end{equation}

\emph{IL.} Off-resonance attenuation in the through path,
\begin{equation}
\text{IL} = -10 \log_{10} (T_\text{through,max}).
\label{eq:IL}
\end{equation}

Resonance wavelengths and linewidths were determined by interpolated curve fitting rather than from the raw wavelength grid. The coupling state (over-coupled, critically coupled, or under-coupled) was assigned by tracking the joint evolution of $Q$, through-port ER, and drop-port ER with gap \cite{Yariv2002}. \rev{The complete set of extracted metrics for all sixteen geometries is tabulated in the Supplementary Document (Tables~S1--S3), together with the full transmission spectra and the individual metric trends (Figures~S1--S26).}

\subsection{Inverse design methodology}
 
To test whether the sweep-derived coupling geometry can be refined, we performed an inverse design study of the bus waveguide in the coupling zone using Tidy3D \cite{Tidy3D}. The three-dimensional model replicated every structural detail of the Lumerical setup ($W_r = 600$~nm, $g = 200$~nm, $R = 10$~$\mu$m, $h = 400$~nm). Because the target window (620--650~nm) lies within the low-$k$ tail of Cs$_3$Cu$_2$Cl$_5$, non-dispersive constants $n = 1.720$ ($k \approx 0$ at 636~nm) for the core and $n = 1.457$ for SiO$_2$ were adopted.
 
The bus waveguide adjacent to the ring was partitioned into three contiguous 1.5~$\mu$m segments, each with an independently variable width bounded to 350--650~nm. Ring shape, gap, and cladding remained frozen. Sensitivity of the drop-port coupling coefficient $\kappa^2$ to each width was estimated by central finite differences ($\Delta = 15$~nm). An Adam optimizer \cite{Kingma2015} with step size 0.015 updated the widths over two successive iterations, followed by broadband validation over 620--650~nm.

Finite-difference gradients were chosen over adjoint sensitivity because the low parameter count (three variables) makes both approaches comparable in cost, and the goal is to benchmark the parametric optimum rather than explore a high-dimensional manifold.

\section{Results and discussion}

\rev{The three subsections that follow present one sweep each, but they are cuts through a single design space rather than separate studies. Each sweep varies one geometric parameter while the other two are held at the baseline values given in Section~2.2, so each isolates one axis of a space whose axes are not independent: the coupling condition that governs $Q$ and ER depends jointly on the mode confinement set by $W_r$, the bus--ring separation set by $g$, and the interaction length set by $R$. Sections~3.1--3.3 therefore establish the individual trends; Section~3.4 assembles them into the trade-off that determines the operating point; and Section~3.5 gives the physical bounds within which that operating point must lie.}

\subsection{Effect of ring waveguide width}

\begin{figure}[t]
\centering
\includegraphics[width=\linewidth]{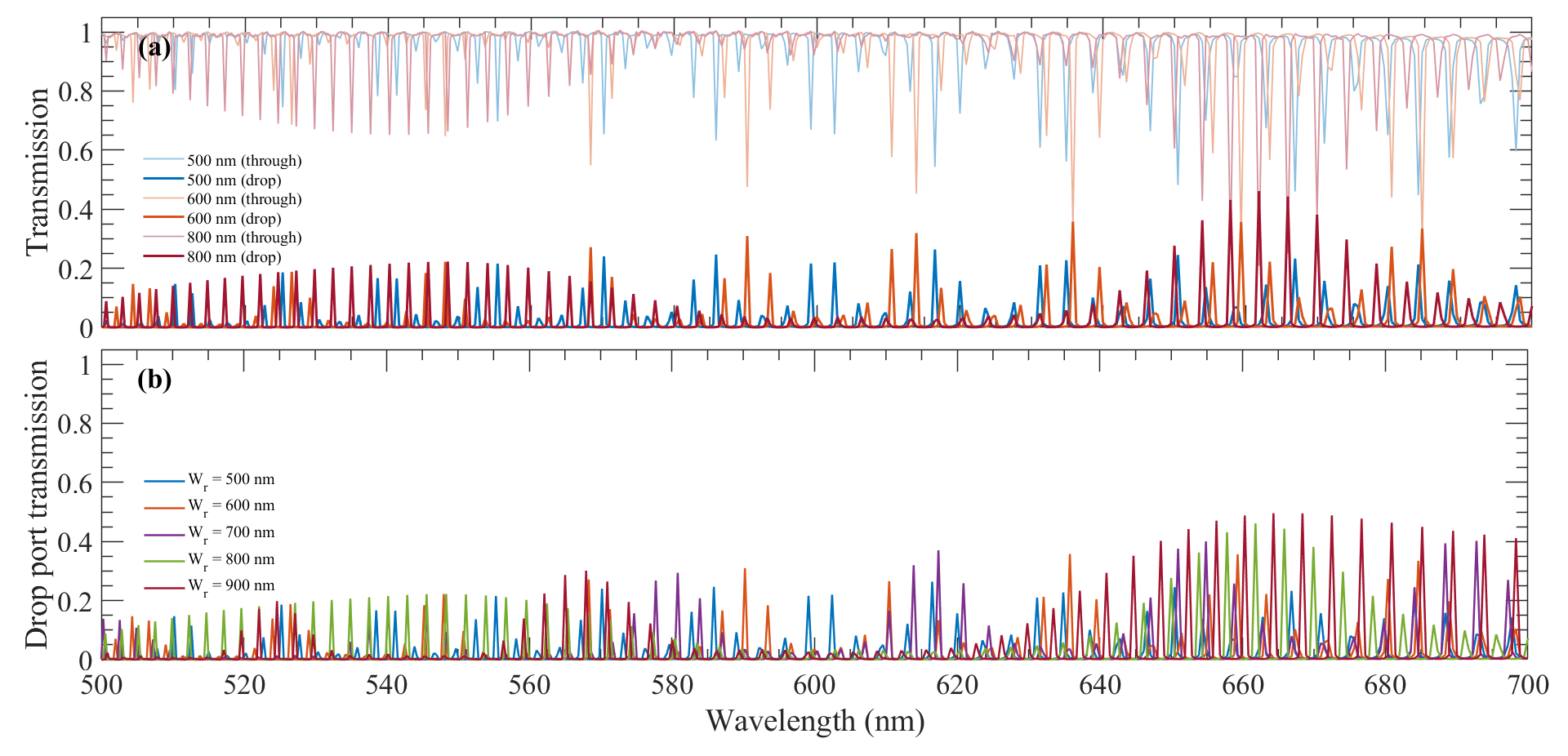}
\caption{(a) Through- and drop-port spectra for $W_r = 500$, 600, and 800~nm at fixed $g = 200$~nm and $R = 10$~$\mu$m. (b) Drop-port response for all five ring widths.}
\label{fig:width_spectra}
\end{figure}

With $g = 200$~nm and $R = 10$~$\mu$m fixed, $W_r$ was stepped from 500 to 900~nm. Representative port spectra appear in Fig.~\ref{fig:width_spectra}. Narrow, high-contrast resonance features are evident for $W_r = 500$ and 600~nm. As the waveguide widens to 800~nm and beyond, the dips broaden and neighboring orders begin to overlap.

\begin{figure}[htbp]
\centering
\includegraphics[width=\linewidth]{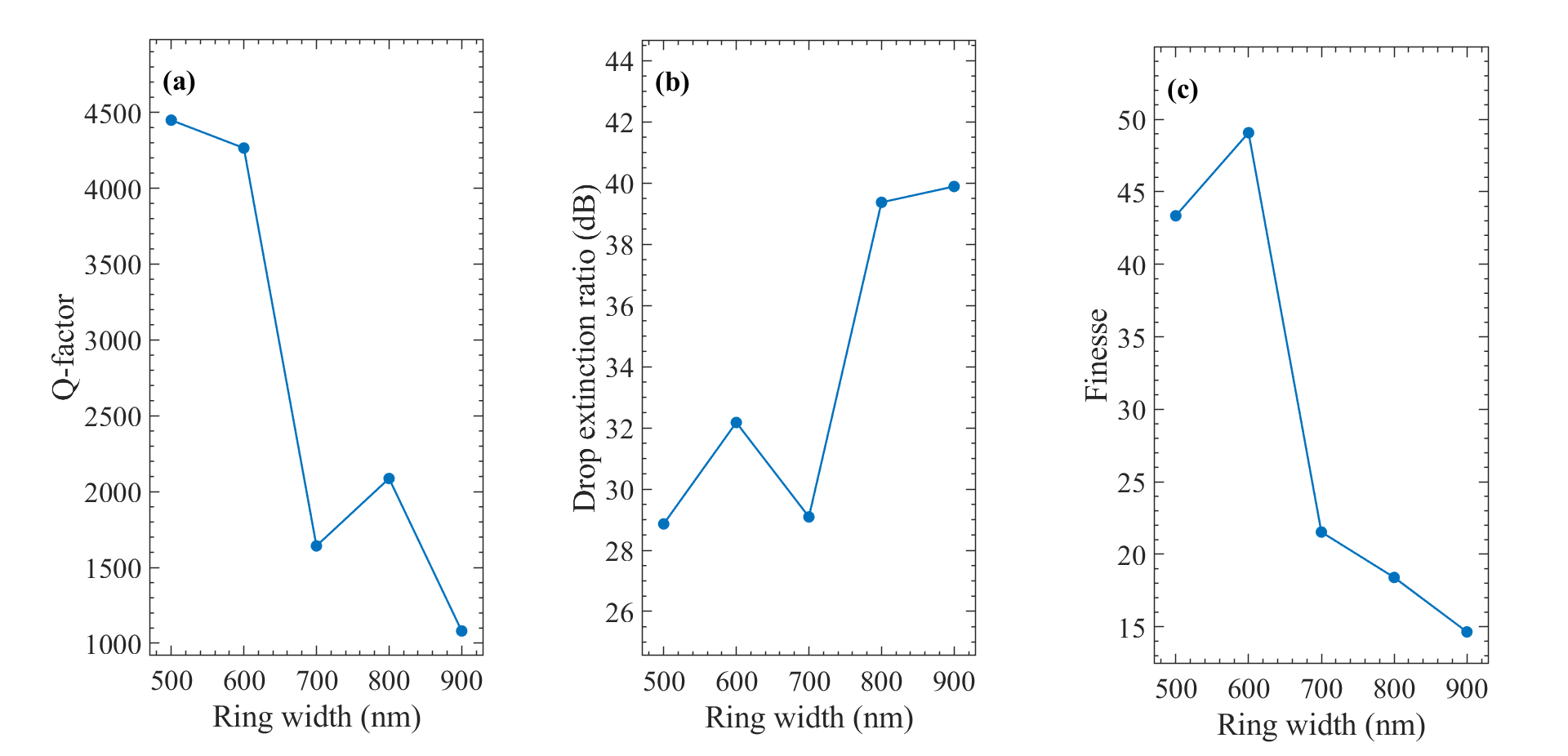}
\caption{Width dependence of (a) $Q$, (b) drop-port ER, and (c) $\mathcal{F}$. $g = 200$~nm and $R = 10$~$\mu$m throughout.}
\label{fig:width_metrics}
\end{figure}

Width-dependent figures of merit are compiled in Fig.~\ref{fig:width_metrics}\rev{, with the full set of extracted values tabulated in Table~S1 and the complete spectra and individual metric trends given in Figures~S1--S8 of the Supplementary Document}. $Q$ falls from ${\sim}4{,}449$ at 500~nm to ${\sim}1{,}081$ at 900~nm, with the steepest decline above 600~nm. A partial recovery at 800~nm does not restore $Q$ to the narrow-width levels. At 600~nm the ring delivers $Q \approx 4{,}265$ alongside 32.2~dB drop-port ER. $\mathcal{F}$ mirrors the $Q$ trend, while drop-port ER rises with width, approaching ${\sim}40$~dB for the two widest guides.

These trends reflect a confinement--coupling trade-off. At $W_r = 500$--600~nm the TE$_0$ mode is tightly bound yet extends far enough into the gap for near-critical coupling \cite{Yariv2002}. Above 700~nm the waveguide begins supporting TE$_1$, introducing inter-modal beating that smears the lineshape \cite{Bogaerts2012, Bahadori2018}. The longer effective interaction arc also increases $\kappa^2$, pushing the device into over-coupling where ER improves at the expense of linewidth \cite{Yariv2002}.

The 500--600~nm window offers the best spectral performance. The 500~nm ring maximizes $Q$; the 600~nm ring provides a better $Q$--ER compromise and preserves single-mode guidance (Section~3.5).

\subsection{Coupling gap dependence and coupling regime transitions}

\begin{figure}[htbp]
\centering
\includegraphics[width=\linewidth]{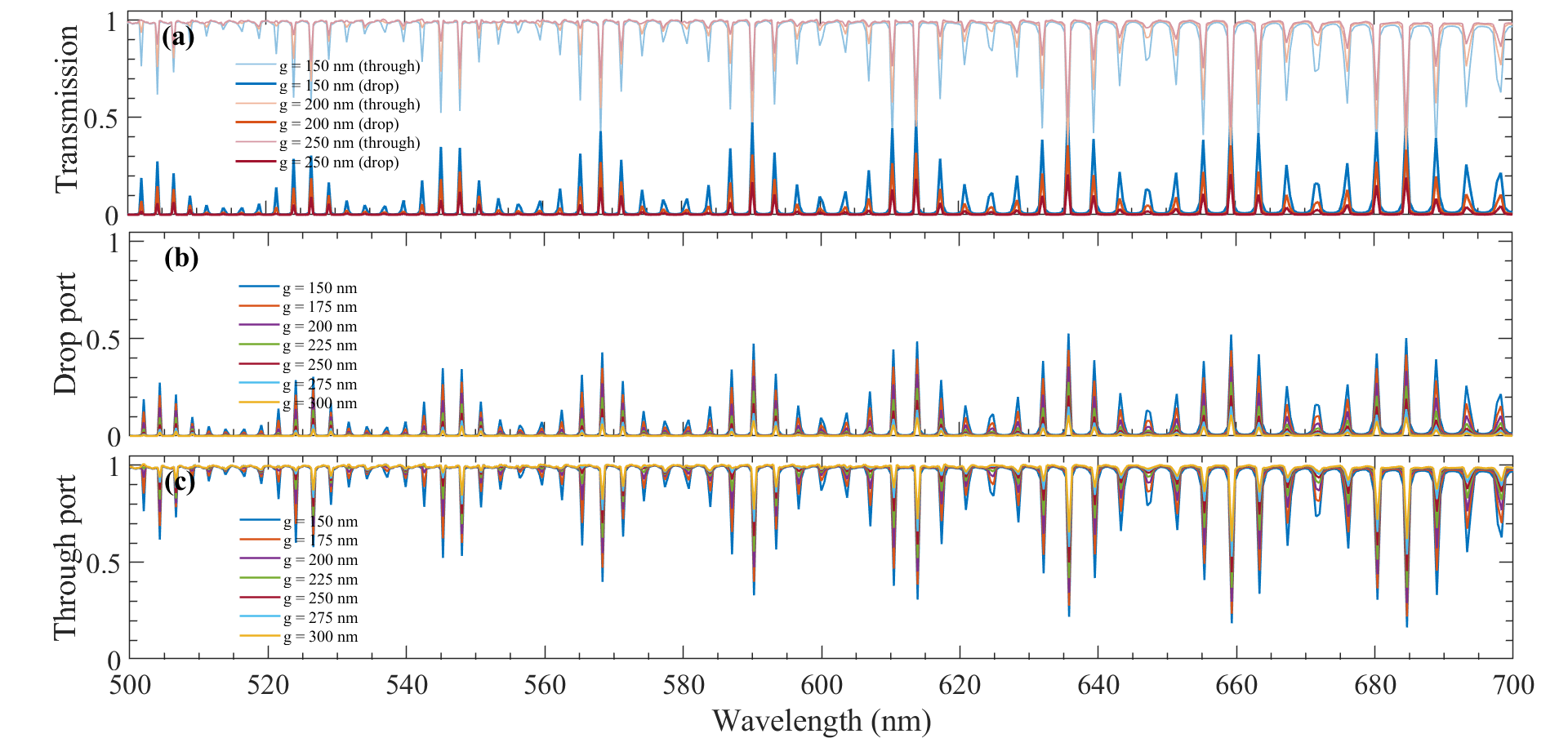}
\caption{(a) Port spectra at $g = 150$, 200, and 250~nm ($W_r = 600$~nm, $R = 10$~$\mu$m). (b) Drop-port and (c) through-port transmission for all gap values.}
\label{fig:gap_spectra}
\end{figure}

With $W_r = 600$~nm and $R = 10$~$\mu$m fixed, $g$ was swept from 150 to 300~nm. Because $g$ sets $\kappa^2$, it directly governs the balance between power injected into the ring and power lost per round trip. Figure~\ref{fig:gap_spectra} displays spectra at three representative gaps.

At $g = 150$~nm the resonances are wide and the drop signal is strong, indicating over-coupling. At 200~nm the features sharpen and the through-port nulls deepen. Beyond 250~nm the drop peaks diminish and through-port dips become shallow, consistent with under-coupling.

\begin{figure}[t]
\centering
\includegraphics[width=\linewidth]{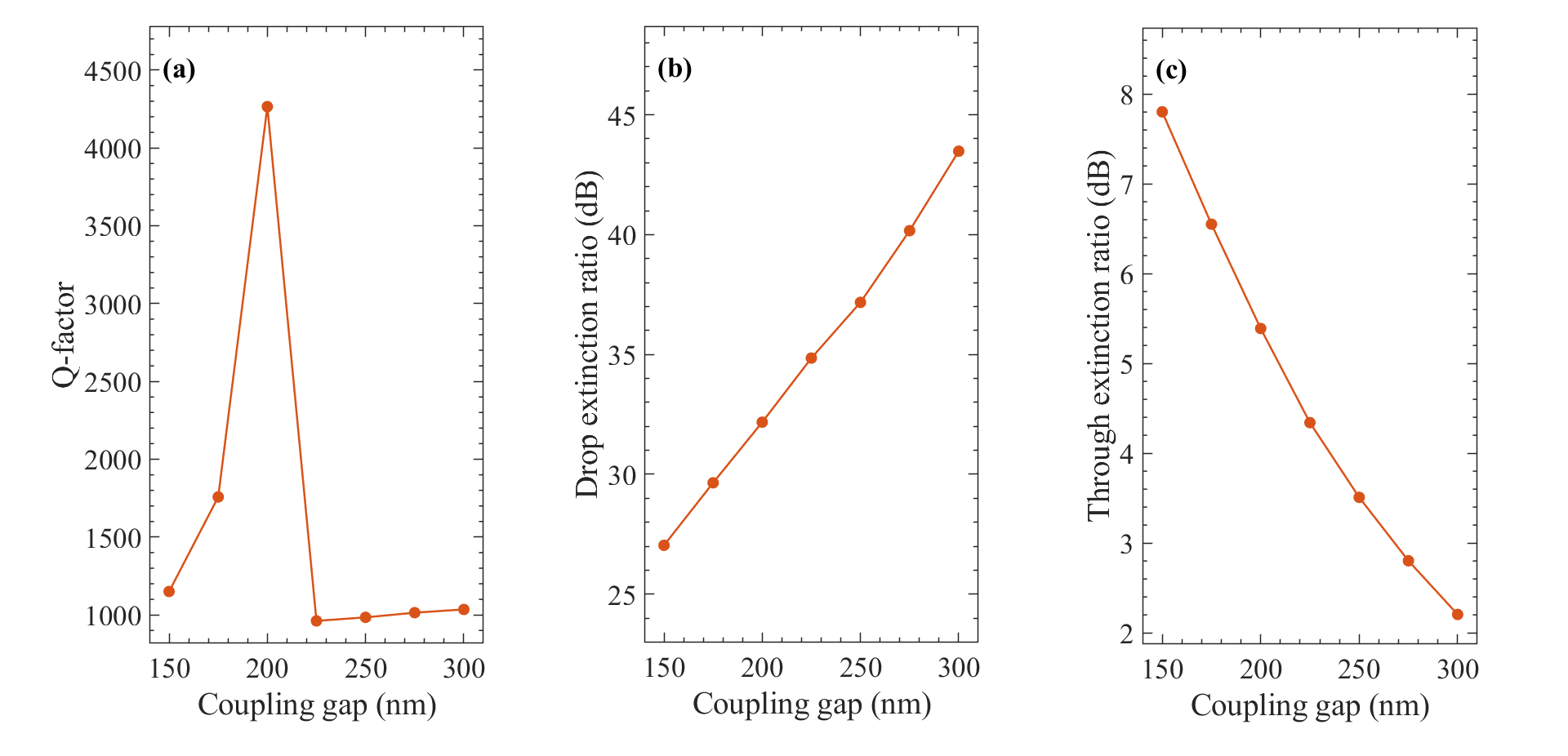}
\caption{Gap dependence of (a) $Q$, (b) drop-port ER, and (c) through-port ER.}
\label{fig:gap_metrics}
\end{figure}

Figure~\ref{fig:gap_metrics} quantifies these trends\rev{; the corresponding numerical values are given in Table~S2, and the complete gap-sweep spectra and metric trends in Figures~S9--S17 of the Supplementary Document}. $Q$ peaks at ${\sim}4{,}265$ for $g = 200$~nm and drops to $\sim 10^3$ on either side. Drop-port ER climbs monotonically with gap; through-port ER follows the opposite slope. A minor wavelength drift at large gaps, caused by gap-dependent perturbation of the ring effective index, was accounted for by extracting all metrics from the same longitudinal mode order.

These observations align with coupled-mode theory for doubly loaded ring cavities \cite{Yariv2002, Rabus2007}. The loaded linewidth depends on the sum of intrinsic round-trip loss and coupling-induced loss at each bus junction. Critical coupling occurs when the two loss channels are equal, producing the deepest through-port null. Below 175~nm excess coupling loss broadens the resonance. Above 225~nm the ring is starved of input power, raising drop-port contrast but curtailing absolute drop throughput \cite{Bogaerts2012}.

The narrow critical-coupling window reflects the modest $\Delta n \approx 0.26$--0.32 between Cs$_3$Cu$_2$Cl$_5$ and SiO$_2$. Platforms with stronger contrast tolerate wider gap ranges before shifting regime \cite{Chrostowski2015}. For this geometry, $g = 200$~nm strikes the best balance between spectral selectivity and power-transfer efficiency. Smaller gaps favor strong extinction but reduce $Q$, while larger gaps curtail drop efficiency despite higher ER contrast.

\subsection{Radius-dependent performance and bending loss threshold}

\begin{figure}[htbp]
\centering
\includegraphics[width=\linewidth]{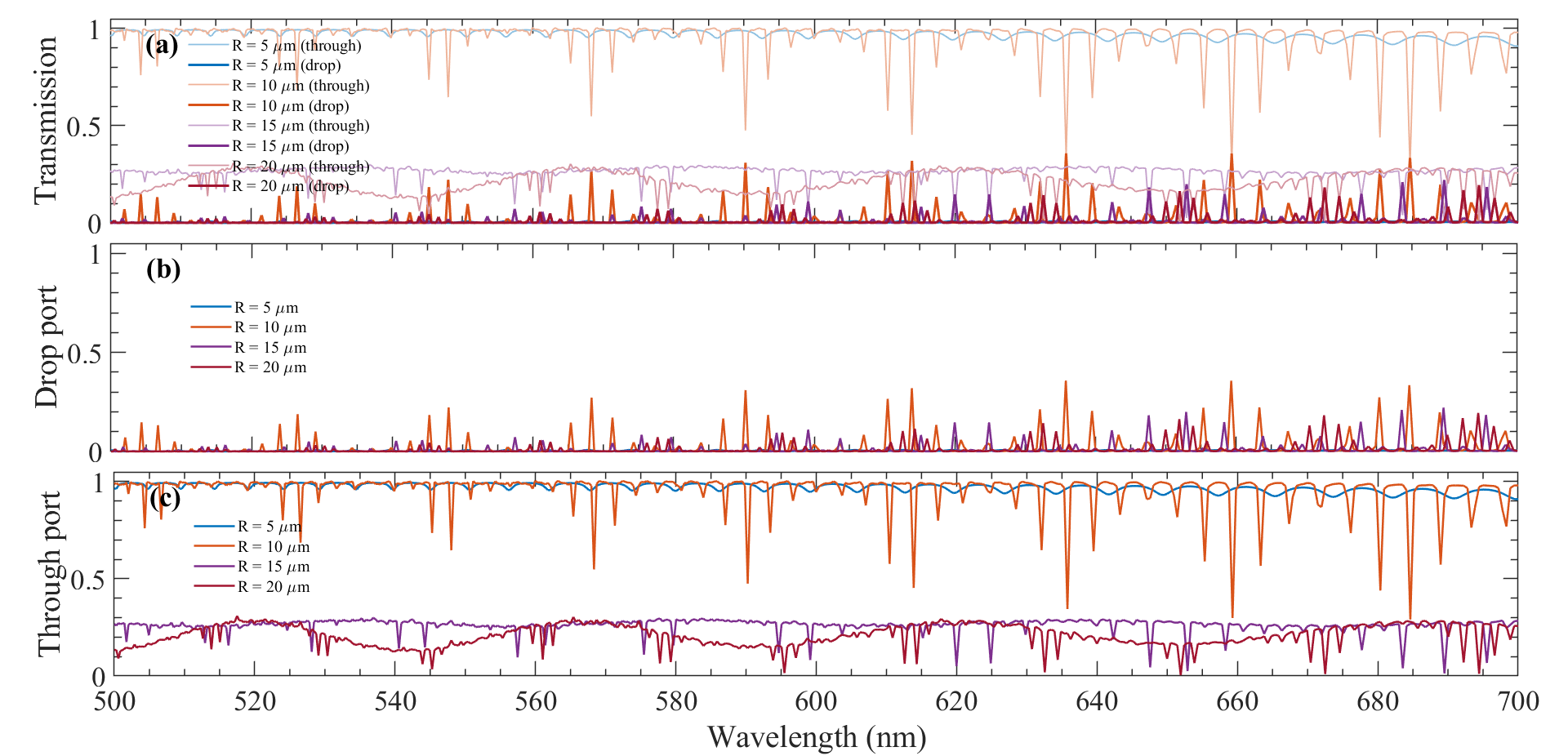}
\caption{(a) Port spectra at $R = 5$, 10, 15, and 20~$\mu$m ($W_r = 600$~nm, $g = 200$~nm). (b) Drop-port and (c) through-port spectra shown separately.}
\label{fig:radius_spectra}
\end{figure}

With $W_r = 600$~nm and $g = 200$~nm held constant, $R$ took the values 5, 10, 15, and 20~$\mu$m. The spectra (Fig.~\ref{fig:radius_spectra}) \rev{show} a \rev{pronounced} performance cliff between 5 and 10~$\mu$m\rev{; extracted values are tabulated in Table~S3 and the corresponding spectra and trends appear in Figures~S18--S26 of the Supplementary Document}.

The 5~$\mu$m ring barely resonates. Its linewidth exceeds 4.7~nm, giving $Q \sim 126$, and nearly all input power bypasses the ring. The mode cannot survive the tight curvature long enough to build up constructive interference. At $R = 10$~$\mu$m the picture changes completely. Sharp resonances appear with $Q \approx 5{,}386$, FSR $= 11.3$~nm, $\mathcal{F} = 95.8$, and negligible IL. This abrupt transition marks the practical minimum bend radius for Cs$_3$Cu$_2$Cl$_5$/SiO$_2$ at visible wavelengths.

At 15 and 20~$\mu$m, FSR narrows as expected from the inverse proportionality to circumference. $Q$ decreases relative to the 10~$\mu$m case and IL rises to ${\sim}5$~dB. Two mechanisms contribute. First, the longer optical path accumulates more propagation loss per round trip. Second, the different curvature alters the effective coupling strength, detuning the device from the critical-coupling condition that was optimized at 10~$\mu$m.

\begin{figure}[t]
\centering
\includegraphics[width=\linewidth]{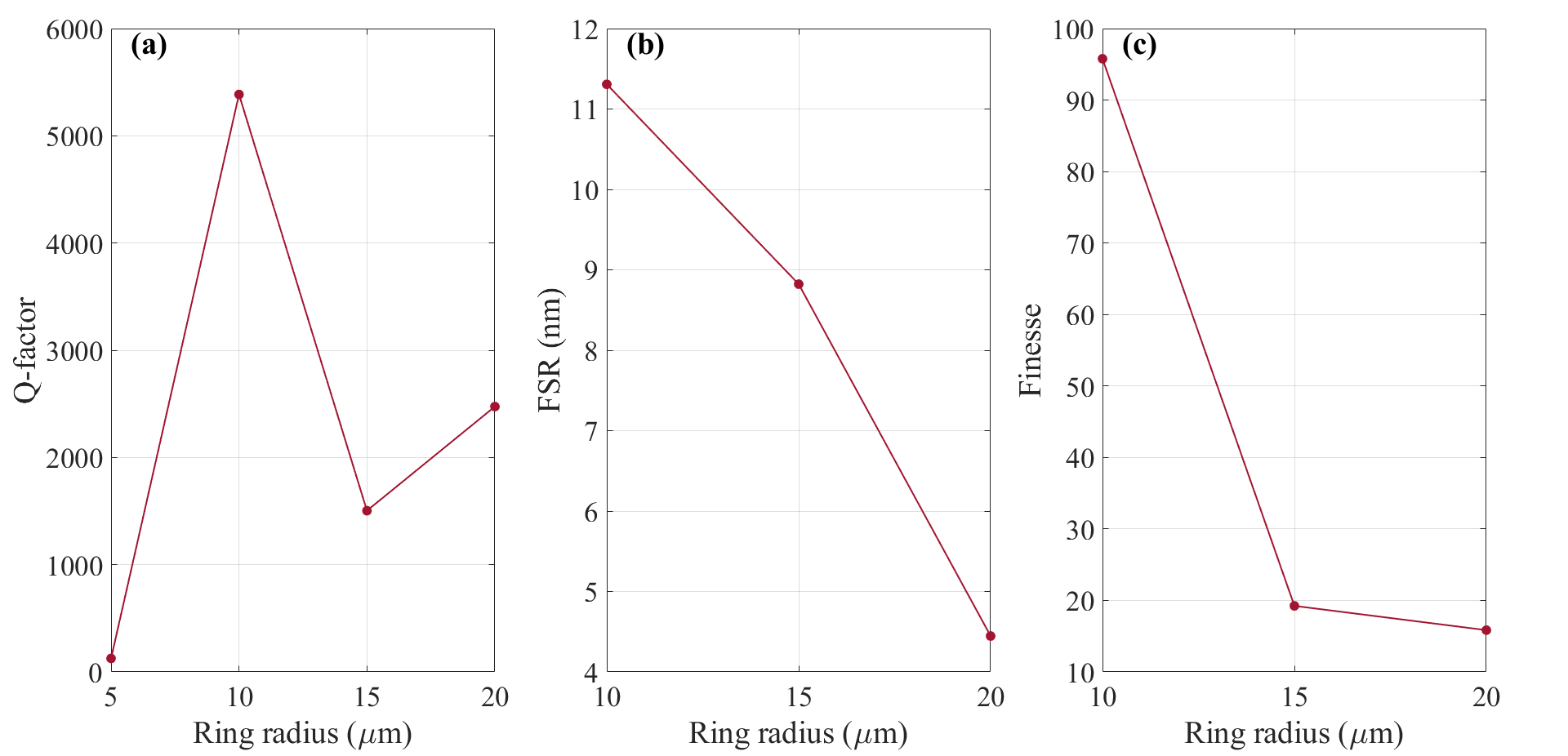}
\caption{Radius dependence of (a) $Q$, (b) FSR, and (c) $\mathcal{F}$. Data for $R = 5$~$\mu$m are omitted in (b) and (c) because no well-defined resonance could be identified.}
\label{fig:radius_metrics}
\end{figure}

Radiation loss from a curved dielectric guide scales exponentially with the inverse of the bend radius \cite{Bogaerts2012}. High-contrast SOI waveguides can operate at few-$\mu$m radii \cite{Bogaerts2012, Xiao2007}. \rev{The weaker confinement of the Cs$_3$Cu$_2$Cl$_5$/SiO$_2$ system pushes the corresponding limit to substantially larger radii. Within the four radii examined here, $R = 5$~$\mu$m is strongly radiation-limited while $R = 10$~$\mu$m is not, so the transition lies between these two values; the sampling used does not locate it more precisely than that.}

Overall, $R = 10$~$\mu$m delivers the best combination of spectral resolution, compact footprint, and manageable propagation loss \rev{among the radii tested}. \rev{The 5~$\mu$m ring is strongly radiation-limited, and the two larger radii increase device area and reduce spectral performance without improving $Q$.}

\begin{figure}[htbp]
\centering
  \includegraphics[width=1.0\textwidth]{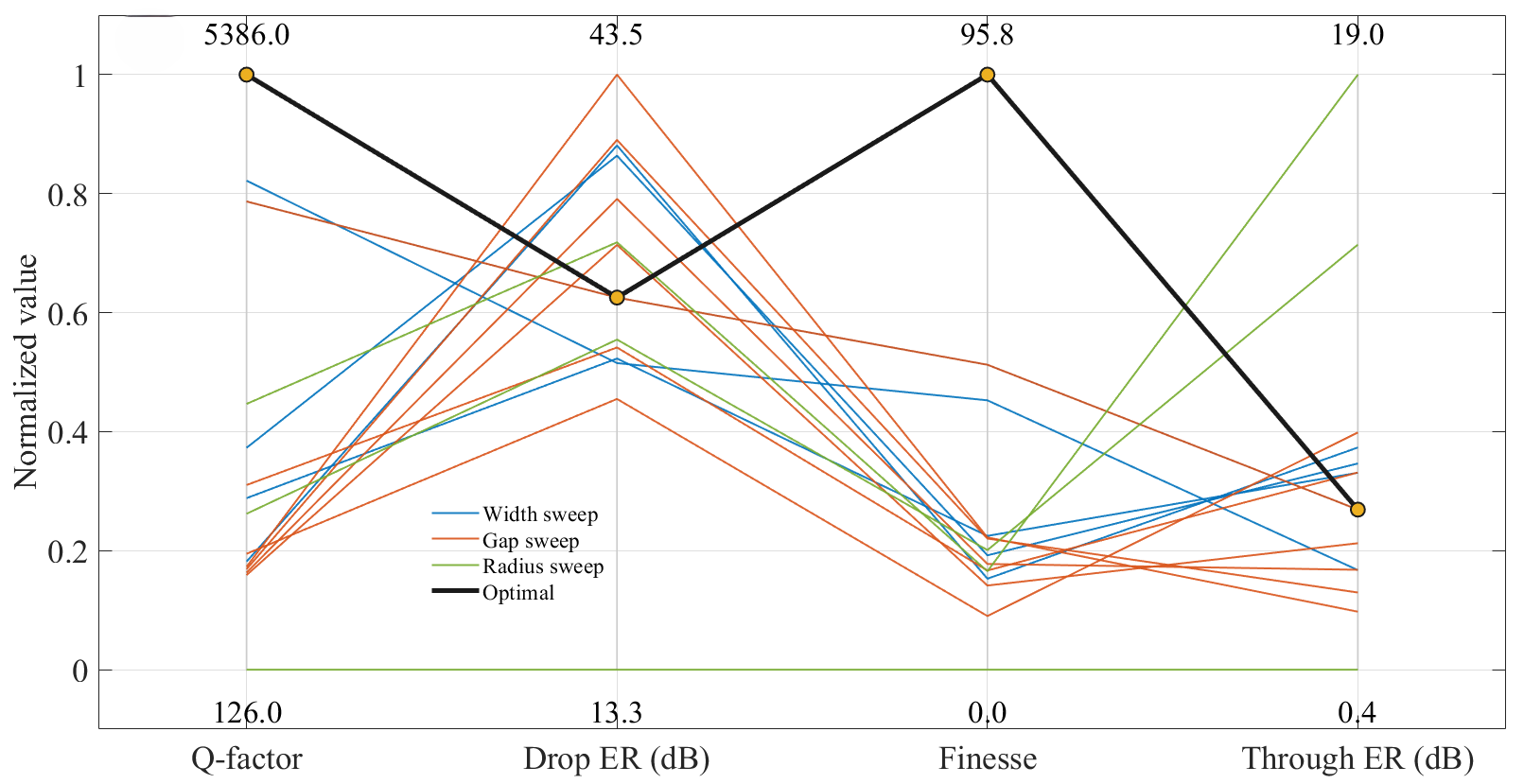}
  \caption{Parallel-coordinates representation of the full parameter-sweep data. Each polyline corresponds to one geometry; axes show normalized $Q$, drop-port ER, $\mathcal{F}$, and through-port ER.}
  \label{fig:design_map_a}
\end{figure}

\subsection{Design trade-off analysis and optimization guidelines}

\begin{figure}[t]
\centering
  \includegraphics[width=0.7\textwidth]{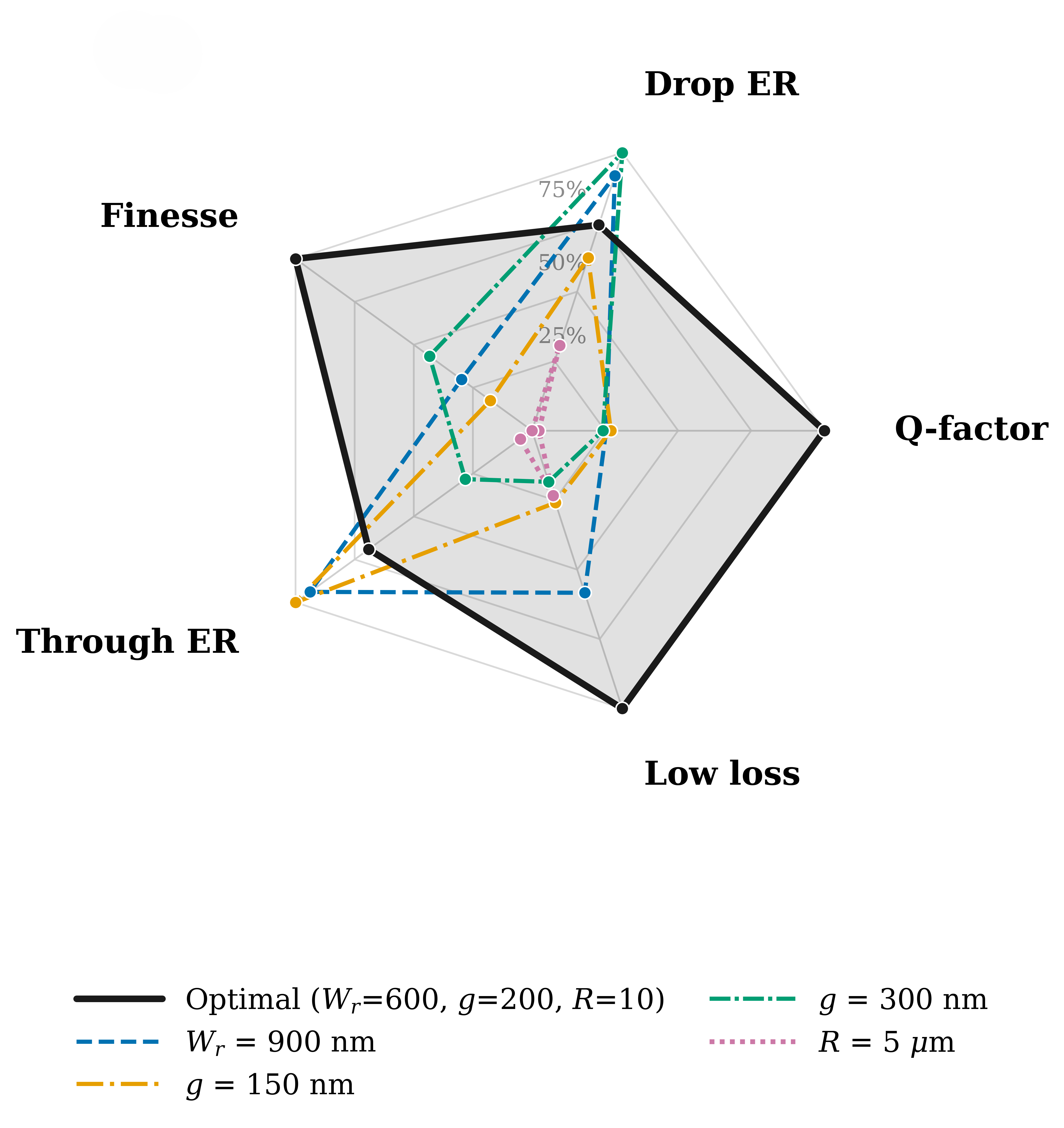}
  \caption{Radar-chart comparison of the optimal design against boundary geometries.}
  \label{fig:design_map_b}
\end{figure}

Figures~\ref{fig:design_map_a} and~\ref{fig:design_map_b} consolidate the width, gap, and radius sweeps into a unified design-space view. In the parallel-coordinates plot (Fig.~\ref{fig:design_map_a}), each polyline represents one tested geometry, and the four vertical axes correspond to $Q$, drop-port ER, $\mathcal{F}$, and through-port ER. The optimal configuration is highlighted. The radar chart (Fig.~\ref{fig:design_map_b}) compares the optimum with boundary cases ($W_r = 900$~nm, $R = 5$~$\mu$m, $g = 150$ and 300~nm) to expose which metrics collapse first when a parameter moves away from its optimal value.

A fundamental $Q$--ER tension pervades the design space. Maximizing $Q$ requires critical coupling, whereas stronger coupling boosts ER at the cost of broader linewidths. This tension is most pronounced in the gap sweep. Moving from $g = 200$ to 150~nm raises drop-port ER by $\sim$5~dB but cuts $Q$ by more than a factor of three. The width sweep exhibits a similar but less abrupt pattern. Widths above 700~nm push ER toward 40~dB, but at the cost of multimode excitation and $Q$ below 2{,}000. The radius sweep adds a third axis of variation. Only $R = 10$~$\mu$m simultaneously achieves high $Q$, high $\mathcal{F}$, and negligible IL. At $R = 5$~$\mu$m the cavity barely resonates, while at 15 and 20~$\mu$m propagation loss degrades $Q$ and IL even though the resonance linewidth itself remains narrow.

The interaction between these three parameters is also important. For example, the critical gap depends on the mode confinement, which is set by $W_r$. Narrower rings confine the mode more weakly, extending the evanescent tail and increasing $\kappa^2$ for a given gap. This means that the 200~nm gap optimum reported here is specific to $W_r = 600$~nm; a wider ring would require a larger gap to reach critical coupling, and vice versa. Similarly, the effective coupling length increases with $R$ because the bus waveguide runs parallel to a longer arc of the ring, which partly explains why the critical-coupling condition shifts when the radius changes.

The selected design ($W_r = 600$~nm, $g = 200$~nm, $R = 10$~$\mu$m) sits at the intersection of the single-mode, near-critical-coupling, and low-bend-loss regions of the parameter space. It achieves $Q \approx 5.4 \times 10^3$, drop-port ER of 32~dB, $\mathcal{F} \approx 96$, and IL $< 0.02$~dB. No other tested geometry simultaneously satisfies all three conditions.

\begin{table}[htbp]
\centering
\caption{Cross-platform comparison of reported MRR performance.}
\label{tab:comparison}
\begin{tabular}{lcccclc}
\toprule
\textbf{Platform} & \textbf{$\lambda$ (nm)} & \textbf{$R$ ($\mu$m)} & \textbf{$Q$} & \textbf{FSR (nm)} & \textbf{Ref.} & \rev{\textbf{Type}} \\
\midrule
SOI                & 1550  & 5      & $2.4 \times 10^5$ & $>$18    & \cite{Xiao2007} & \rev{Expt.} \\
Si$_3$N$_4$        & 780   & 23     & $3.7 \times 10^6$ & ---      & \cite{Ji2017} & \rev{Expt.} \\
TFLN               & 1550  & 80     & $1.2 \times 10^6$ & ---      & \cite{Zhang2017LN} & \rev{Expt.} \\
Polymer            & 770   & 100    & $5 \times 10^5$   & $\sim$0.6 & \cite{Lin2024} & \rev{Expt.} \\
\textbf{Cs$_3$Cu$_2$Cl$_5$} & \textbf{500} & \textbf{10} & $\mathbf{5.4 \times 10^3}$ & \textbf{11.3} & \textbf{This work} & \rev{\textbf{Sim.}} \\
\bottomrule
\end{tabular}
\end{table}

Table~\ref{tab:comparison} places these results alongside established MRR platforms. The absolute $Q$ is several orders of magnitude below SOI or Si$_3$N$_4$ devices \cite{Xiao2007}, which benefit from far higher $\Delta n$ and decades of process optimization. Our simulations include material absorption but omit fabrication-related scattering, so the reported $Q$ should be interpreted as an upper bound for a given geometry. Still, $Q \sim 5 \times 10^3$ is adequate for visible-band wavelength filtering and evanescent-field sensing, where the figure of merit depends on $Q$ divided by mode volume rather than on $Q$ alone \cite{Chrostowski2015}. The distinguishing advantage of Cs$_3$Cu$_2$Cl$_5$ is that it provides a non-toxic waveguide material transparent where crystalline Si is opaque, filling a gap in the material toolbox for visible integrated photonics.

\rev{A comparison against prior \emph{studies}, rather than against reported devices, is also worth making, since it is the study design and not the absolute $Q$ that establishes what is new here. Computational optimization of microring geometry is itself well established: Dizaji and Habibiyan, for example, coupled Bayesian optimization to physics-based constraints to select microring geometry for squeezed-light generation \cite{Dizaji2025SR}. Work placing halide perovskites into integrated photonic circuits has likewise been extensive, but has drawn on the perovskite for an active optical function rather than using it as the passive guiding medium \cite{Wang2021AOM, Han2024NC}. Studies of Cs$_3$Cu$_2$Cl$_5$ in particular have remained at the material level, reporting electronic structure, optical constants and luminescence without addressing guided-wave behavior \cite{Ali2023Heliyon, Zhang2020}. The present study differs from all three groups in treating a lead-free copper halide as the passive core, taking its optical constants from first principles rather than from a fitted dispersion model, and mapping the resulting design space in the visible band. The comparison in Table~\ref{tab:comparison} should therefore be read as situating a first-pass computational design against mature experimental platforms, not as a like-for-like performance claim.}

Recommended operating windows are summarized in Table~\ref{tab:guidelines}. These guidelines can be adapted depending on the target application. Sensor designs benefit from near-critical coupling to maximize $Q$ and hence the wavelength-shift sensitivity per unit change in cladding $n$. Filter designs that prioritize extinction contrast over spectral selectivity can tolerate slightly wider gaps (e.g.\ $g = 225$--250~nm) at the expense of lower $Q$. For applications requiring both high $Q$ and compact footprint, \rev{$R = 10$~$\mu$m is the smallest radius examined here that retains the required performance}; reducing $R$ further would require either a higher-$n$ cladding or a thicker waveguide to improve mode confinement\rev{, and would need its own radius sweep to establish where the practical limit falls}.

\begin{table}[htbp]
\centering
\caption{Design guidelines for Cs$_3$Cu$_2$Cl$_5$ add-drop MRRs in the visible band.}
\label{tab:guidelines}
\begin{tabularx}{\linewidth}{@{}lc>{\raggedright\arraybackslash}X@{}}
\toprule
\textbf{Parameter} & \textbf{Value} & \textbf{Rationale} \\
\midrule
$W_r$ & 500--600~nm & Ensures fundamental-mode guidance and $Q > 4{,}000$. Broader rings suffer TE$_1$ excitation and $Q$ degradation. \\
\addlinespace
$g$ & $\sim$200~nm & Targets critical coupling for balanced $Q$ and ER. Tighter gaps over-couple; wider gaps starve the drop port. \\
\addlinespace
$R$ & $\geq$10~$\mu$m & \rev{Smallest radius tested that avoids the strong radiation loss seen at 5~$\mu$m}. Yields peak $Q$ and a practical 11~nm FSR. \\
\bottomrule
\end{tabularx}
\end{table}

\begin{figure}[t]
\centering
\includegraphics[width=0.8\linewidth]{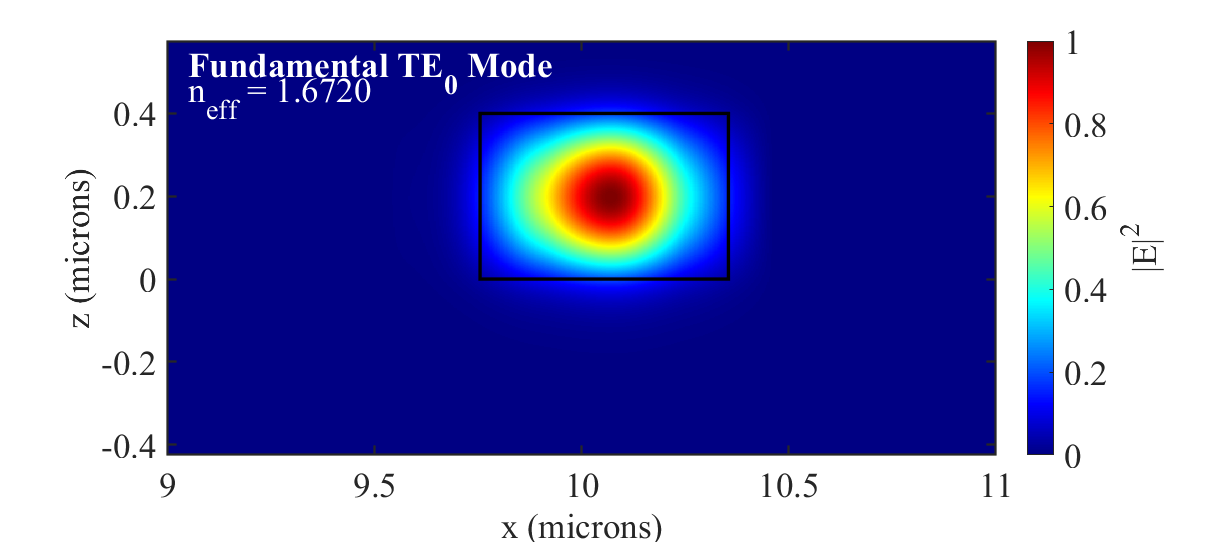}
\caption{Normalized $|E|^2$ of the TE$_0$ mode ($W_r = 600$~nm, $h = 400$~nm). $n_\text{eff} = 1.6720$; the black rectangle marks the core boundary.}
\label{fig:mode_profile}
\end{figure}

\rev{\subsection{Physical limits of the design space}}

\rev{The sweeps locate an operating point, but they do not by themselves say what bounds the space that point sits in. Four limits do, and stating them in closed form allows the rules developed here to be transferred to a different geometry, film thickness or related compound without repeating the simulations.}

\rev{\emph{Lower bound on bend radius.} Radiation loss from a curved dielectric guide grows exponentially as the radius falls, with a decay constant set by the index contrast between core and cladding \cite{Bogaerts2012, Rabus2007}. The bound is therefore not a sharp cutoff but a threshold that moves with $\Delta n$, and the modest contrast of the Cs$_3$Cu$_2$Cl$_5$/SiO$_2$ system ($\Delta n \approx 0.26$--0.32 across the operating window) places it at considerably larger radii than a high-contrast platform such as SOI, which tolerates radii several times smaller \cite{Xiao2007}. The present sweep brackets rather than resolves this threshold: it lies between the 5~$\mu$m ring, which is strongly radiation-limited, and the 10~$\mu$m ring, which is not. Raising the film thickness or introducing a higher-index cladding would move it down.}

\rev{\emph{Upper bound on ring width.} The width is bounded above by the onset of the first higher-order guided mode, beyond which inter-modal beating degrades the lineshape irrespective of the coupling condition. The eigenmode solve reported in Section~3.6 finds no TE$_1$ solution at $W_r = 600$~nm and a TE$_1$ solution at $W_r = 700$~nm; on the 100~nm sampling used, the onset therefore lies between these two widths. This is the ceiling reflected in the design guidelines of Table~\ref{tab:guidelines}, and locating it more sharply would require a finer width sweep of the cross-section.}

\rev{\emph{Two-sided bound on the coupling gap.} The bus--ring power coupling coefficient falls approximately exponentially with gap, $\kappa^2 \propto \exp(-2\gamma g)$, where $\gamma$ is the transverse decay constant of the evanescent field in the cladding. Critical coupling requires $\kappa^2$ to match the round-trip loss, so the admissible gap window is narrow when the loss is small and narrows further as the evanescent tail shortens. This is why the critical gap reported here is specific to $W_r = 600$~nm: a wider ring confines the mode more tightly, shortens the tail, and requires a smaller gap to reach the same $\kappa^2$.}

\rev{\emph{Material bound on the achievable $Q$.} Absorption in the core imposes a wavelength-dependent practical limit on the intrinsic quality factor that no choice of geometry can remove. For a mode with group index $n_g$ and confinement factor $\Gamma$ in a medium of extinction coefficient $k$, the propagation loss is $\alpha = 4\pi k/\lambda$ and the absorption-limited quality factor reduces to}
\begin{equation}
\rev{Q_\text{abs} = \frac{2\pi n_g}{\Gamma \alpha \lambda} = \frac{n_g}{2\,\Gamma k}.}
\label{eq:Qabs}
\end{equation}
\rev{The wavelength dependence therefore enters entirely through $k(\lambda)$. Because $k$ falls steeply through the short-wavelength end of the operating window (Fig.~\ref{fig:nk_plot}(b)), this limit relaxes by orders of magnitude across the visible band, and is most restrictive at the blue edge. A designer targeting the highest attainable $Q$ on this platform should therefore operate as far toward the red end of the transparency window as the application permits, and should evaluate Eq.~\eqref{eq:Qabs} at the intended wavelength rather than assuming a single figure across the band.}

\rev{\emph{Temperature and pressure.} As noted in Section~2.1, the optical constants used throughout derive from a static ground-state calculation at zero applied pressure, without explicit finite-temperature effects. Finite operating temperature and applied pressure may modify the resonance through thermal expansion and thermo-optic effects. Neither was modeled here, and we are not in a position to quantify them for this compound: to our knowledge the thermo-optic coefficient and the thermal expansion tensor of Cs$_3$Cu$_2$Cl$_5$ have not been reported. The results presented here should therefore be read as a nominal ground-state, zero-pressure design map. The temperature dependence of the resonance is a limitation of the present study rather than something the present calculations bound, and evaluating it would require either a thermo-optic calculation or experimental characterization.}

\subsection{Mode profile verification}

Single-mode operation was confirmed by solving for the guided eigenmodes of the ring cross-section ($W_r = 600$~nm, $h = 400$~nm) in Lumerical MODE. The TE$_0$ field map (Fig.~\ref{fig:mode_profile}) shows a well-confined intensity maximum centered within the Cs$_3$Cu$_2$Cl$_5$ core. $n_\text{eff} = 1.6720$, falling between the core ($n \approx 1.78$) and cladding ($n = 1.46$), as expected for a guided mode.

The evanescent field penetrates roughly 200--300~nm into the SiO$_2$ on each side. This is consistent with the strong gap sensitivity observed in Section~3.2 and suggests that the platform is well suited to evanescent-field $n$ sensing, where changes in the cladding $n$ shift the resonance wavelength. At $W_r = 600$~nm the confinement is sufficient to suppress higher-order guided modes, supporting the single-mode assumption used throughout the analysis. The eigenmode solver detects a TE$_1$ solution only for $W_r \geq 700$~nm, corroborating the $Q$ degradation in Section~3.1.

\subsection{Inverse design optimization and cross-platform validation}
 
\begin{figure}[htbp]
\centering
\includegraphics[width=\linewidth]{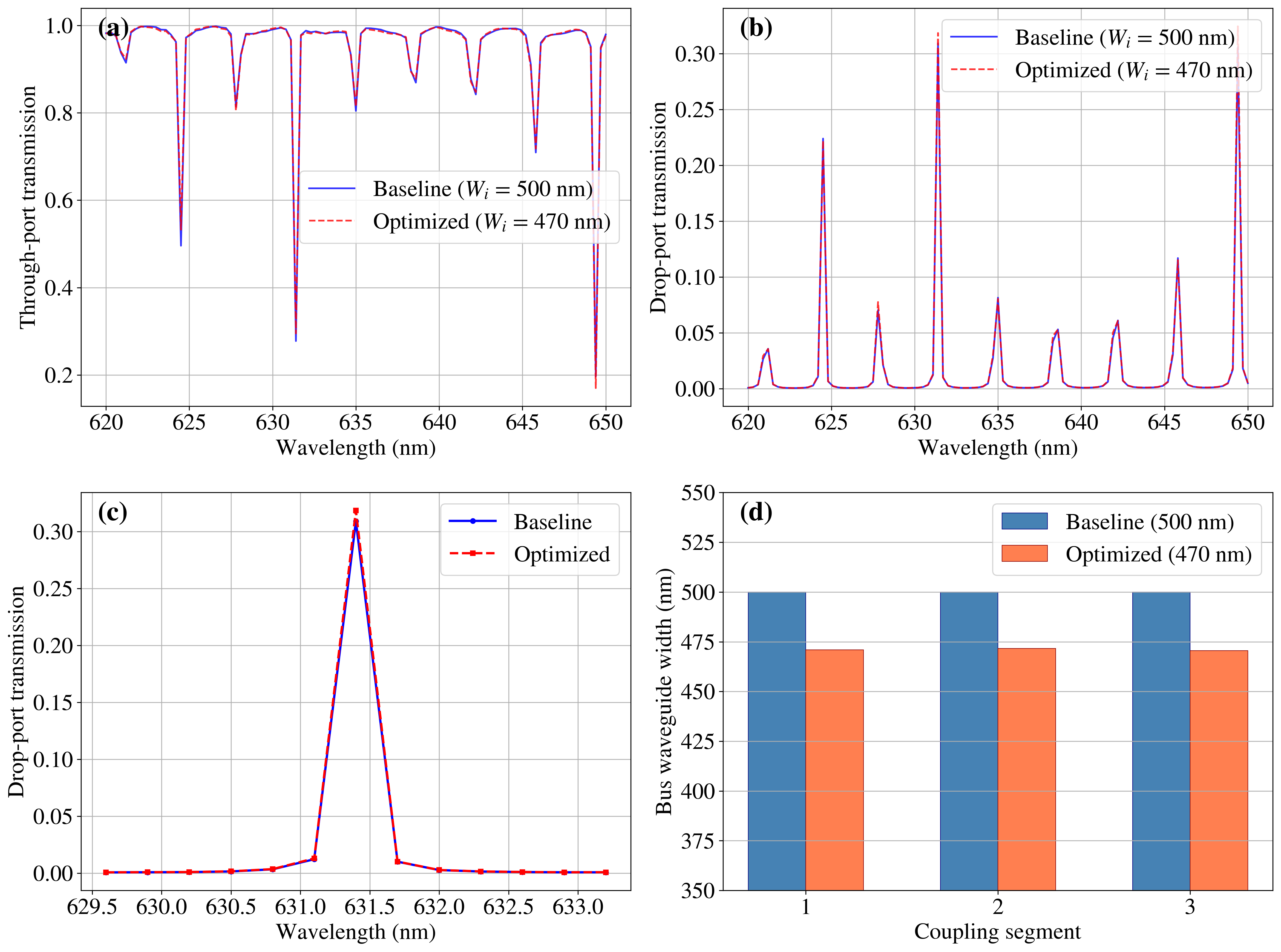}
\caption{Coupling-region inverse design. (a,\,b) Through- and drop-port spectra for baseline ($W_i = 500$~nm) and optimized ($W_i = 470$~nm) bus widths. (c) Magnified view near $\lambda_\text{res} = 631.4$~nm showing the 3\% drop-coupling gain. (d) Optimized width profile across the three segments.}
\label{fig:invdes}
\end{figure}

After two gradient-update steps the optimizer converged to a nearly uniform bus width of ${\sim}470$~nm across all three segments, a 6\% narrowing relative to the 500~nm baseline. The convergence trajectory and final width profile are in Fig.~\ref{fig:invdes}(c--d).
 
Figures~\ref{fig:invdes}(a--b) overlay the broadband through- and drop-port spectra from Tidy3D for both layouts over the 620--650~nm validation window. The signatures (sharp through-port dips paired with narrow drop-port peaks) closely reproduce the Lumerical predictions from Sections~3.1--3.3. This provides solver-independent corroboration of both the device model and the DFT-derived $n$ and $k$.
 
At the resonance near 631.4~nm (Fig.~\ref{fig:invdes}(c)), the optimized layout raises peak drop-port transmission from 0.309 to 0.319, a 3.0\% relative improvement, without shifting $\lambda_\text{res}$, broadening the FWHM, or degrading $Q$. Table~\ref{tab:invdes} compares detailed metrics. Through-port ER decreases by 0.27~dB; drop-port ER is unchanged. All three optimized widths collapse to a single value, indicating that the parametric sweep had already placed the geometry near a local coupling-efficiency maximum.
 
\begin{table}[t]
\centering
\caption{Baseline vs.\ inverse-designed coupling-region performance.}
\label{tab:invdes}
\begin{tabular}{lccc}
\toprule
\textbf{Metric} & \textbf{Baseline} & \textbf{Inv.\ Design} & \textbf{$\Delta$} \\
\midrule
$W_i$ (nm)                 & 500    & 470    & $-30$   \\
$\lambda_\text{res}$ (nm)  & 631.40 & 631.40 & ---    \\
Peak drop $T$              & 0.3093 & 0.3185 & $+$0.009 \\
FWHM (nm)                  & 0.311  & 0.311  & ---   \\
$Q$                        & $\sim$5{,}386 & $\sim$5{,}386 & ---   \\
ER$_\text{through}$ (dB)   & 5.56   & 5.29   & $-$0.27  \\
ER$_\text{drop}$ (dB)      & 26.79  & 26.81  & $+$0.02  \\
\bottomrule
\end{tabular}
\end{table}

Two conclusions follow. First, the Lumerical--Tidy3D spectral agreement shows that the DFT-derived constants and the simulation workflow are reproducible across independent solvers. Second, the marginal uplift from inverse design supports the guidelines of Table~\ref{tab:guidelines}. The combination $W_r = 600$~nm, $g = 200$~nm, $R = 10$~$\mu$m constitutes a well-optimized operating point for the Cs$_3$Cu$_2$Cl$_5$ platform.

It should be noted that the present inverse design explores only bus width as a free parameter within a limited coupling zone. A richer formulation, such as pixel-level topology optimization of the full coupling region using adjoint gradients, could access non-intuitive geometries with potentially larger performance gains. Such an extension is deferred to future work.

\begin{table}[htbp]
\centering
\caption{\rev{Simulated} sensitivity of the optimized MRR to \rev{representative} geometric \rev{tolerances}.}
\label{tab:sensitivity}
\begin{tabular}{lcccc}
\toprule
\textbf{Parameter} & \textbf{Error} & $|\Delta Q|$ & $|\Delta$ER$_\text{drop}|$ (dB) & $|\Delta\mathcal{F}|$ \\
\midrule
$W_r$ & $\pm$50 nm      & 702 & 0.06 & 5.5 \\
$g$   & $\pm$25 nm      & 398 & 2.60 & 1.2 \\
$R$   & $\pm$0.5 $\mu$m &  69 & 0.84 & 7.7 \\
\bottomrule
\end{tabular}
\end{table}

\subsection{Fabrication feasibility}

While the present study is purely computational, recent Cs$_3$Cu$_2$Cl$_5$ thin-film advances lend credibility to the proposed designs. Wang et al.\ reported scalable aqueous synthesis with 97\% \rev{photoluminescence quantum yield (PLQY)}, depositing large-area scintillator coatings by \rev{blade coating} \cite{Wang2023AOM}. Zhou et al.\ obtained dense films via hot-injection on glass (72.4\% PLQY) \cite{Zhou2022}. \rev{Sequential vacuum evaporation has yielded flexible films of approximately 100~cm$^2$} \cite{Qiu2024}, and aerosol-assisted CVD has produced quartz-supported layers with PLQY $\sim$99\% \cite{Vega2019}. These diverse routes \rev{indicate} that the waveguide dimensions identified here are \rev{compatible with} current thin-film technology\rev{. No devices were fabricated in the present study; the discussion here is intended only to establish that the dimensions proposed are not out of reach of reported deposition methods}.

Beyond film deposition, the practical viability of the proposed MRR also depends on how tightly the geometric parameters must be controlled during lithographic patterning. Table~\ref{tab:sensitivity} summarizes the impact of representative fabrication errors on the three principal figures of merit at the optimal design point. $Q$ is most sensitive to ring width, with a $\pm$50~nm error producing a shift of $\sim$700. The coupling gap dominates the ER sensitivity, where $\pm$25~nm translates to a 2.6~dB change in drop-port contrast. $\mathcal{F}$ is most affected by radius variation. Notably, the radius tolerance ($\pm$0.5~$\mu$m) is the easiest to satisfy with standard electron-beam or deep-UV lithography, while the tighter gap tolerance ($\pm$25~nm) may require process calibration on the specific Cs$_3$Cu$_2$Cl$_5$/SiO$_2$ stack.

\section{Conclusion}
 
\rev{We have mapped, by simulation, the geometry--performance landscape of add-drop MRRs based on the lead-free visible-band halide Cs$_3$Cu$_2$Cl$_5$. DFT-computed $n(\lambda)$ and $k(\lambda)$ were imported into 2.5D varFDTD simulations for the parametric scans of $W_r$, $g$ and $R$, and the selected geometry was subsequently re-solved with full three-dimensional FDTD.}
 
Ring widths of 500--600~nm sustain the highest $Q$ while maintaining single-mode operation. The gap sweep traces a textbook over-coupled $\to$ critically coupled $\to$ under-coupled progression, with the optimum near $g = 200$~nm. A sharp bending-loss threshold appears between 5 and 10~$\mu$m. The 10~$\mu$m ring yields peak $Q$, whereas smaller radii are radiation-limited and larger radii accumulate excess propagation loss.
 
Combining all sweeps, the geometry $W_r = 600$~nm, $g = 200$~nm, $R = 10$~$\mu$m offers the most balanced trade-off. Tidy3D corroborates the predicted spectral response. Inverse design of the coupling zone, executed by segmenting $W_i$ into three freely adjustable sections, achieves only a 3\% gain in drop-port coupling. All three segments converge to 470~nm, \rev{which supports the interpretation} that the parametric sweep had already located a near-optimal coupling geometry.
 
These results \rev{provide quantitative design rules for Cs$_3$Cu$_2$Cl$_5$ MRRs, which to our knowledge have not previously been reported, and support the case for lead-free visible-wavelength integrated photonics as a direction warranting experimental investigation}. Future work will include experimental demonstration, investigation of related copper-halide compositions (Cs$_3$Cu$_2$Br$_5$, Cs$_3$Cu$_2$I$_5$), and deployment of adjoint-based topology optimization over pixel-level design regions to explore higher-dimensional design spaces that the segmented-width approach of this work cannot access.

\begin{backmatter}

\bmsection{Acknowledgment}
The authors acknowledge the computational facilities provided by the Department of Electrical and Electronic Engineering (EEE), BUET, throughout the duration of this work.

\bmsection{Disclosures}
The authors declare no conflicts of interest.

\bmsection{Data Availability Statement}
Representative FDTD simulation scripts, processed datasets, and supporting materials are openly available at the GitHub repository:
\url{https://github.com/debnath-shoumik/Cs3Cu2Cl5-MRR-HighQ-DesignRules}.
Additional data may be obtained from the corresponding author upon reasonable request.

\end{backmatter}

\bibliography{references}

\end{document}